# On a Finite Universe with no Beginning or End

Peter Lynds[1]

c/- 61B Rimu St, Waikanae, Kapiti Coast 6010, New Zealand

Based on the conjecture that rather than the second law of thermodynamics inevitably be breached as matter approaches a big crunch or a black hole singularity, the order of events should reverse, a model of the universe that resolves a number of longstanding problems and paradoxes in cosmology is presented. A universe that has no beginning (and no need for one), no ending, but yet is finite, is without singularities, precludes time travel, in which events are neither determined by initial or final conditions, and problems such as why the universe has a low entropy past, or conditions at the big bang appear to be so "special," require no causal explanation, is the result. This model also has some profound philosophical implications.

## 1. Introduction

In his monumental 1781 Treatise, *Critique of Pure Reason* [1], the German philosopher, Immanuel Kant, argued that the notions of the universe having no beginning and stretching back infinitely in time, versus it beginning at some finite time in the past, were equally contradictory. If the universe had no beginning, there would be an infinite period of time before any event. To Kant, this idea – that time could stretch back forever – was absurd. On the other hand, if the universe did have a beginning, what happened beforehand to cause it? And what before that? He thought this second notion, that the universe could have a finite past, was equally contradictory. Although we have since realized that it is meaningless to talk of the concepts of time and space without the presence of matter (as would be the case before a big bang singularity at the said beginning of the universe),[2] this still does not actually address Kant's paradox, nor does it provide any answer to the most basic and fundamental question of cosmology: what *caused* the big bang and where did the matter that became this unrivalled explosion come from? Although Kant's argument has been given great attention over the two centuries since, I do not think that it has been fully realized just what a perfect paradox he put forward here. Few would disagree that it is suggesting something deep to us about our assumptions concerning time, cause and cosmology, but I do not think that the actual root of this contradiction has yet been fully questioned i.e., what is going wrong with our assumptions that this great breakdown in reason occurs?

Based on the conjecture that rather than the second law of thermodynamics inevitably be breached as matter approaches a big crunch or a black hole singularity, the order of events should reverse, in this paper, a model of the universe which provides a solution to the mystery of its origin, as well as to the paradox posed by Kant, is presented. Implications of the model, one which appears to be compatible with general relativity and standard hot big bang theory, include that the universe has neither a beginning nor ending, but yet is finite; singularities are never actually encountered in nature; events are neither determined by initial or final conditions; problems such as why the universe has a low entropy past [2, 3], or conditions at the big bang appear to be so "special" [4], require no *causal* explanation; and that time travel is not possible. Several predictions made by the model are then explained, followed by an assessment of its specific weaknesses and strengths. This paper also has some profound philosophical implications, a discussion of which will largely be left for a more suitable forum. Lastly, I should note that this paper is largely non-mathematical, as, with one notable exception, I do not feel that the arguments herein can be made so, while, for reasons that should become evident, filling in the model with the relevant equations is somewhat arbitrary. As the paper deals with some difficult and unfamiliar concepts, but I would still like to try to have it be readily understandable, I have tried to write it in as clear, simple and explanatory way possible, including devoting much of the initial sections to providing introduction and context. For these reasons, this is a somewhat non-typical and awkward paper – it sitting somewhere between physics and philosophy, and having too much of each to be either. I am hopeful that readers will be sympathetic of this.

---

[1] E-mail: peterlynds@xtra.co.nz
[2] Unless one were specifically referring to intervals of time and space as "represented" by clocks or rulers, strictly speaking, it would be meaningless to talk about time and space *after* the big bang also.

## 2. The Second Law of Thermodynamics and Time Reversal

As most readers will be aware, the second law of thermodynamics is related to the fact that heat can never pass spontaneously, or of itself, from a hot to hotter body. Hot flows to cold. As a result of this, natural processes that involve energy transfer tend to have one direction and to be irreversible. This law also predicts that the entropy of an isolated system increases with time, with entropy being the measure of the disorder or randomness of energy and matter within a system. This is shown by Ludwig Boltzmann's great entropy formula, $S = k \log V$, where k is Boltzmann's constant $1.3806505(24) \times 10^{-23}$ J K$^{-1}$, and log V is the natural logarithm of the number of possible microstates corresponding to the macroscopic state of a system [5]. Because of the second law of thermodynamics, both energy and matter in the universe are becoming less useful, or more disordered, as time goes on. As the second law is thought to give a one-way direction in time to events, and all of the laws of physics, with the exception of the second law of thermodynamics, are time symmetric and can equally be reversed,[3] it is also sometimes referred to (somewhat misleadingly) as the "arrow of time." The second law of thermodynamics, however, is of course a statistical claim. Chance fluctuations and decreases in entropy on an atomic scale are reasonably commonplace. Although it is generally believed that the only truly isolated system in nature is the universe itself, and as a consequence, it is a safe bet to say that the second law of thermodynamics will never be breached just by chance, as the statistical possibility cannot ever be absolutely discounted, when strictly stating the second law of thermodynamics, it is more correct to say that the entropy of an isolated system is *extremely likely* to increase over time, rather than always.[4]

---

[3] The second law of thermodynamics is not reversible, for if it were, a system would evolve in the direction in which entropy was generally decreasing, something that, by definition, the second law statistically forbids. Although the question of how such an asymmetry can arise from completely time-symmetric mechanical laws (this being augmented by Loschmidt's reversibility paradox [6, 7]) is a very pertinent one, it does not make the second law any less irreversible. It is important to note, however, that the second law makes no distinction which direction in time entropy should increase, whether towards (what we call) the future, or (what we call) the past. This point seems to have been first recognized by Boltzmann himself [5]. Surprisingly, what does not seem to have been realized is the reason why entropy does not *simultaneously* increase in both time directions (and in this sense, the increase of entropy is asymmetric). It is unable to, for if a system is evolving with entropy increasing in one time direction (for example, towards what we call the future), and then went to evolve, with entropy increasing, in the opposite time direction (towards what we call the past), as the system's past is already set, entropy would be decreasing in that direction. A past with entropy increasing would not be an option, as a system cannot have a different past if it already has one. As such, a system can evolve only in one direction, with entropy generally increasing, at a time, whether that direction be towards what we call the future, or what we call the past.

In a 1985 paper [8], Stephen Hawking discussed what effect the second law of thermodynamics might have on the future of the universe. Like Thomas Gold in the 1960s [9, 10], he had come to the conclusion that if the universe began to contract, something he believed would eventually happen, the thermodynamic arrow of time would reverse with it. Everything would go into the reverse of the way we experience things today: light would travel back to the stars, and broken eggs on the floor would miraculously put themselves back together again. Unlike Gold, who thought that when the universe began to contract and grow smaller, hot could not flow to cold, but would tend to hotter objects, Hawking came to this conclusion by way of quantum cosmology and the idea that when the universe began to contract, entropy would also decrease because the universe would have to return to a symmetrical state of low entropy and high order at a big crunch [8, 11].

This meant that Hawking and Gold's "big crunch" would be what we call the "big bang," and for events to get there, everything in our current epoch of expansion would have to play over, backwards, as the universe recontracted. This was in spite of the consideration that for events to reverse in the way Hawking and Gold had envisaged, radiation would need to be converging back towards the stars of such a contracting phase from within the universe's present epoch of expansion. In other words, light might be returning from our eyes to the stars today. Such a situation would give rise to some very troublesome problems, including a system having two arrows of time pointing in opposing directions. Inevitable interaction of opposing time-directed radiation would mean that at least some of the time-reversed radiation would have a different evolution to that radiation's (previously) non time-reversed evolution; its evolution could not actually be symmetrically time-reversed. The question of what might cause the radiation to begin reversing, while the rest of the system continued to evolve as usual, is also problematic. During the universe's recontraction, effects would precede causes, a person would grow younger as time passed, and one would be dead before they were born. Nevertheless, as the same thermodynamic processes would underlie our physiology and brain processes, people living during the recontraction would also *think* in the direction in which entropy was decreasing. Because this would mean that their mental processes, memory and direction of causality would also be reversed, everything would seem quite normal to such back-to-front people.

The theory, however, was dependant on something of a big ask – that entropy would decrease as the universe began to contract, thus violating the second law of thermodynamics in the process. If one grants that this can happen, one must also face up to all the apparent contradictions and science fiction-like possibilities that such a situation would naturally entail. Hawking eventually changed his mind about this idea when he was shown that entropy would still go on increasing after the universe began to contract, and called it his "greatest mistake"[12]. Puzzlingly, while rightfully pointing out the error of assigning a preferred direction in time to events in physics, and also tackling Hawking on a "double standard" in his cosmology – a case of Hawking deriving a temporally asymmetric consequence from a symmetric theory – in his book, *Time's Arrow and Archimedes' Point*, Huw Price still claimed that the arrow of time should flip with the universe's contraction [3].

---

[4] For good discussions of the second law of thermodynamics, see, for example: Zeh, H. D., *The Physical Basis of the Direction of Time,* Springer-Verlag, Berlin, 1989; Davies, P. C. W., 1974 *The Physics of Time Asymmetry,* University of California Press Berkeley and Los Angeles, 1974; and, Penrose, R., *The Road to Reality, A Complete Guide to the Laws of the Universe,* Ch. 27, Knopf, 2005.

## 3. The Big Crunch and Cyclic Universes

Probably like millions of others, when I first read Hawking's book, *A Brief History of Time* [11], it was also the first time that I gave more than just a little thought to the origins, and possible future, of the universe. Central to my interest was that I thought it made much more sense for the universe to be positively spatially curved (i.e. Omega$_M$ = $\rho_0$ / $\rho_c$ = > 1). If the universe had begun with a big bang and then expanded, it seemed equally plausible that one day it could do the reverse and contract to a big crunch. There is an obvious symmetry to such a situation. Furthermore, rather than being billions or trillions the amount, it seemed an odd coincidence that the amount of mass required to form a closed universe was so relatively close to what could be detected (according to 2003 WMAP results, approximately 27%, with Omega$_M$ = 0.27, inclusive of dark matter [13]; while by combining data from the WMAP study and surveys such as SDSS [Sloan Digital Sky Survey] and SDF [Two-Degree Field], 2004 results suggest another two-thirds of the universe's mass to be made up of an unknown type of energy named "dark energy" [14]). Most crucially, however, an eternally expanding cosmos did not help understand the birth of the universe any better. Indeed, it only added to the mystery of what could have possibly happened to *cause* the big bang.

If the universe were positively spatially curved and destined for a big crunch, this naturally posed the question of what might happen next. There seemed to be two general options: either the universe would contract to a singularity, a point of infinite density and geometric space-time curvature, and everything would cease to be; or alternatively, it might bounce back with a great explosion. This "big bounce" would be much like, or possibly exactly the same as, the big bang before it. If the latter, and the universe had exactly the same configuration as the previous big bang, not only would the explosion be exactly the same, but so too would the entire evolution of the universe following it. Such cyclic ideas, whether the universe is thought to be exactly the same in each cycle, or completely different, are not new. Before the Christian view of history took hold, where things are always seen to take place in a linear fashion and only once, cyclic views of the world and cosmos were the prevalent belief in ancient times [15]. More recently, however, in the 1930s, Richard Tolman also pondered the possibility of a cyclic cosmos, and questioned what would happen if a closed universe were to survive its contraction to a big crunch, and burst forth again in a new cycle [16]. Unfortunately, as Tolman realized, the universe would gather entropy during each new phase, and to make up for it, it would have to grow each time like a runaway snowball. This made its continued progression from one cycle to the next unfeasible. As the universe would be bigger and bigger in each new cycle, it also meant that, looking backwards in time, such a universe would need to have had a beginning [17].

In 2002, however, Paul Steinhardt and Neil Turok put forward a similar idea with what they called the cyclic or "ekpyrotic" model [18] – a "Pre-big bang" model of the universe that shares a number of common features with another first developed by Gabriele Veneziano in the early 90s [19, 20]. According to Steinhardt and Turok and their work with M-theory, which states that space-time has 11 dimensions, our universe may be just in the current cycle of an infinite number of previous ones. They further assert that our four-dimensional "brane" is moving among the remaining dimensions or branes, which are hidden at very small or very large lengths. They argue that a big crunch/big bang bounce occurs when two such branes collide. They also claim that the density of matter is completely finite during such a collision, and that a singularity only occurs in the sense that the dimension that separated these branes disappears momentarily during the collision (recently, a number of other models involving big bounces have appeared which use String or Loop quantum gravity theory to propose similar big crunch singularity avoidance mechanisms). Steinhardt and Turok's model relies upon "dark energy" [14] – a type of repulsive gravitation, which was proposed to explain recent observations that indicate that the universe is expanding at an ever-accelerating rate. In their model, dark energy is required to reduce entropy while the universe expands, thus avoiding Tolman's conclusion that each new cycle would gather more and more entropy [16, 17, 21].[5]

## 4. Reversing an Assumption About Time Reversal

When I first began to think seriously about these questions, I found it strange that it seemed little consideration had been put into how the second law of thermodynamics might affect the big crunch itself.[6] Namely, as the universe contracted and became smaller and smaller, it would also become hotter and hotter. As it neared a big crunch, gravity would become so strong, and the universe so small, that nothing could escape. It would be much like hot air inside a balloon, that when contracted, becomes more and more dense. At a certain critical point, when the balloon gets too small and the air inside too dense and hot, it would explode, allowing the hot air inside to flow to and merge with the cooler air outside. What might happen to the universe at this point though? There would be a big difference between the situation with the balloon, because outside of the rapidly contracting universe, there would not actually be anything there. The whole system (the contracting universe) would be completely self-contained and isolated. Moreover, as the universe became smaller, hotter, and gravity stronger and stronger, nothing – including heat – could escape. Hot could not flow to cold. As everything continued to become more and more condensed, heat could only flow, of itself, from hot to hotter. This would suggest a violation of the second law of thermodynamics.

On a microscopic level, this would presumably mean that, as the universe exponentially compressed and heated, hot particles would more and more often begin interacting with hotter particles (rather than cooler ones). As the system compressed further and particle collisions, energies, and particle and photon density also increased further, when photons were emitted by particles and *later* absorbed by other particles, the secondary particles would more and more often absorb photons (heat) in the meantime. As the system compressed further still, a point would be reached where the particle and photon density, and the rate of photon absorption, was such that – more often than not – the secondary particles would be hotter than the initial particles. The balance of the system's heat would begin flowing from hot to hotter. (For convenience sake and ease of explanation, for the remainder of this paper, the word entropy will be used to indicate heat flow. An increase in entropy thus represents heat spontaneously flowing from hot to cooler, and a decrease in entropy, hot to hotter. Talk of the second law of thermodynamics will be within this context).

---

[5] There have been several other models proposed that, in various ways, could be classed as being "cyclic." These include those by Gödel, K., An Example of a New Type of Cosmological Solutions of Einstein's Field Equations of Gravitation, *Rev. Mod. Phys.* Vol. 21, p. 447, 1949; Linde, A., Eternally Existing Self-Reproducing Chaotic Inflationary Universe, *Phys. Lett. B* 175, 395, 1986; Gott, R. J. III, and Li, L., Can the Universe Create Itself? *Phys. Rev. D*, Volume 58, Issue 2, 1988; Israel, W., and Sikkema, A. E., Black-hole mergers and mass inflation in a bouncing universe, *Nature,* 349,45, 1991; and, Smolin, L., *The Life of the Cosmos*, Oxford University Press, USA, 1999. As they have little direct relation to the arguments of the present paper, however, I will avoid going into them.

[6] There are exceptions to this. See footnote 7, Penrose (1979), and Markov (1984).

The natural question then became, what would happen if the second law of thermodynamics were breached? People such as Hawking [8, 11] and Gold [9, 10] had assumed that all physical processes would go into reverse. In other words, they had assumed that events would take place in the direction in which entropy was decreasing, rather than increasing as we observe today. Furthermore, they had assumed that entropy would decrease in the direction in which the universe contracted towards a big crunch (in their case, towards what we call the big bang). But if the second law correctly holds, on a large scale, entropy should still always increase. Indeed, what marks it out so much from the other laws of physics in the first place, is that it is asymmetric – it is not reversible. If all of the laws of physics, with the exception of the second law of thermodynamics, are time symmetric and can equally be reversed, it became apparent that if faced with a situation where entropy might be forced to decrease rather than increase, *rather than actually doing so*, the order of events should simply reverse, so that the order in which they took place would still be in the direction in which entropy was increasing. The second law would continue to hold, events would remain continuous, and no other law of physics would be contravened.

## 5.  A Finite Universe with no Boundaries

This conclusion was then applied to a possible big crunch. What might happen if the second law of thermodynamics were to be breached as the universe became extremely contracted, dense and hot?[7] As outlined earlier, rather than entropy decrease and the second law not hold, the order of events would simply reverse so that events took place in the direction in which entropy was still increasing. In relation to the big crunch, this would mean that at the point where the second law *would have* been breached, any future events in this direction (where entropy would be decreasing) would reverse. Because it would represent the first opportunity at which events could take place in a direction in which entropy was increasing, this would mean that events would immediately take up at where the big crunch singularity *would have* been had events not reversed, and in this direction, no singularity would be encountered. The universe would then expand from where the big crunch singularity would have been had events not reversed (i.e. the big crunch reversed), and with events going in this direction, entropy would still be increasing, no singularity would be encountered, and no laws of physics would be contravened. They would all still hold.

---

[7] The possibility of the second law of thermodynamics being violated and the arrow of time resultantly flipping (ala Hawking and Gold) as events approached a singularity, has been considered (see Penrose, R., Singularities and Time-Asymmetry, pp. 612, in *General Relativity: An Einstein Centenary Survey*, Eds. Hawking, S. W. and Israel, W., Cambridge University Press, 1979; and, Markov, M. A., Problems Of A Perpetually Oscillating Universe, *Annals Phys*. 155:333-357, 1984), being generally dismissed on the grounds of seeming contradiction and non-physicality.

Such a situation could only last so long. Very quickly, events would reach the point at which the second law would have originally been breached had events not reversed. In the current direction, the universe would face an order of events (the entire history of the universe, stretching back to the big bang) in which, in this direction, entropy would be decreasing. If the second law were to still hold and entropy still be able to increase, the order of events would be reversed here once more, immediately taking up again at the first chance they could in the direction in which entropy was increasing. This would be the big bang.

At this point, it becomes apparent that this would not only lead things back to the big bang, but it would actually *cause* it. The universe would then expand, cool, and eventually our solar system would take shape. It would also mean that this would be the exact repeat of the universe we live in now. Something further becomes evident, however, and it is perhaps the most important (and will probably be the most misunderstood and puzzled over) feature of this model. If one asks the question, what caused the big bang?, the answer here is the big crunch. This is strange enough. But is the big crunch in the past or the future of the big bang? It could equally be said to be either. Likewise, is the big bang in the past or future of the big crunch? Again, it could equally be said to be either. The differentiation between past and future becomes completely meaningless. Moreover, one is now faced with a universe that has neither a beginning nor end in time, but yet is also finite and *needs no beginning*. The finite vs. infinite paradox of Kant completely disappears.

Although if viewed from our normal conception of past and future (where we make a differentiation), the universe would repeat over and over an infinite number of times, and could also be said to have done so in the past. Crucially, however, if one thinks about what is actually happening in respect to time, no universe is in the future or past of another one. It is exactly the same version, once, and it is non-cyclic. If so desired, one might also picture the situation as an infinite number of the same universe repeating at exactly the same time. But again, if properly taking into account what is happening in respect to time, in actuality, there is no infinite number of universes. It is *one and the same*.

To perhaps help picture the situation, imagine that the period of time between the big bang and a subsequent big crunch is a mere 10 seconds. A clock begins at $t = 0$ at the big bang, the universe then expands, and at 5 seconds, it then begins to contract. If it were to contract to a big crunch singularity, it would reach the singularity at exactly $t = 10$ seconds. However, just before a singularity inevitably results, at 9.9 seconds events reverse, so the clock immediately goes from 9.9 seconds to 10 seconds without the intervening interval of 0.1 seconds passing. The universe then expands from where the big crunch singularity *would have* been had events not reversed, and the interval between 10 seconds and 9.9 seconds passes. At 9.9 seconds, the universe reaches the point where events originally reversed as the universe contracted towards a big crunch, and is now faced with an order of events, with entropy decreasing, stretching all the way back to the big bang singularity. Subsequently, the order of events are reversed once more, and the clock immediately restarts (inasmuch as the word "restarts" can have meaning here) at $t = 0$, at the big bang, with no singularity being encountered. That is to say, for example, that if one uses the Friedmann-Lemaître-Robertson-Walker (FLRW) metric to describe the evolution of the universe, one can see that, as the evolving equation of the scale factor, $a(t)$, converges to 0 if we work back to the big bang, if we treat the opposite direction of evolution, the order of events in which entropy is increasing, the scale factor will be diverging from $a(t) = 0$. Moreover, in this direction of evolution, the time interval will *always* be diverging from $t = 0$, rather than ever converging to it; $t$ will never go to 0 (or in the case of a big crunch, go to 10). Of course, the same can also be said for any physical magnitude, mass and temperature included, going to infinity at $t = 0$ or $t = 10$. Furthermore, as such singular points do not represent intervals, and a $t$ value must, when the order of events reverse in this model, $t$ will never be (or *at* in the context of some imagined "instant" [22, 23]) 0 at the big bang, or 10 at the big crunch; it will be some interval, however small.

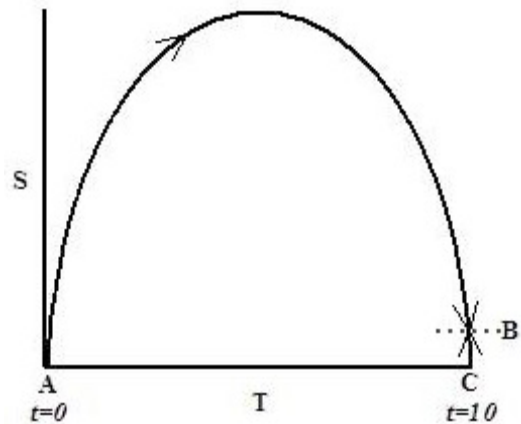

[Figure 1]

**A.** Universe expands from the big bang.[8]
**B.** Point at which, as the universe approaches a big crunch (C), the second law of thermodynamics would be breached in direction B→C.
**C→B.** Order of events reverse so entropy can continue increasing. Universe expands from big crunch (reversed) with no singularity being encountered, very quickly reaching original point at which the second law of thermodynamics would have been breached had events not reversed.
**A→B.** Faces order of events in which entropy is decreasing (B→A). Rather than second law be breached, order of events reverse once more, with no big bang singularity being encountered.

It is important to note that when the clock restarts at the big bang, the universe is not in the future or past of another one. In a sense, it is time itself that restarts (although, again, nothing in fact actually "restarts"), so there is no past or future universe. Because of this, no conservation laws are violated. It is also important to note that it is simply just the order of events that reverse – something that would be immediate. Time does not begin "flowing" backwards to the big bang, nor does anything travel into the future or past of anything, including time and some imagined "present moment" [22, 23]. Indeed, this model contains another interesting consequence. As there is no differentiation between past and future in it, and, strictly speaking, no event could ever be said to be in the future or past of another one, it would appear to provide a clear reason as to why time travel is not possible. In relation to future and past, there is clearly nothing *there* to travel into. Physically speaking, the same can be said for travel *through* an interval of time, a flow of it, as well as space-time [22, 23].

Time travel is precluded due to a further reason arising from this model. Probably the greatest problem with time travel as a theory is that, due to it having already happened, the past seems set and unchangeable, and if it were possible to travel into it, one would then be faced with a number of impossible situations, including the well-known grandfather paradox [24, 25]. But what about the future? As it is seen as not yet having happened, it did not seem that time travel into the future would be quite as problematic. However, if the past and future are completely interchangeable (as they are in this model), and what we call the past is set, then what we call the future would also be set. In other words, time travel into the future would not be possible either.

---

[8] Note that this diagram illustrates the universe's expansion as decelerating rather than accelerating, as currently appears to be the case. The universe being closed and eventually collapsing, however, is the relevant factor here.

Based on the understanding provided by this model, it is now also interesting to note that if the second law of thermodynamics were to be violated if the universe began to contract (as was argued by Hawking and Gold, etc), rather than the universe's current epoch of expansion playing over, backwards, as events returned to the big bang, the whole contraction of the universe to the big crunch would be reversed – at the culmination of which, events stretching back to the big bang would also be reversed. That is, the only change to the present model would be that point B (see Figure 1) would be at the universe's transition from expansion to contraction, rather than just before a big crunch.

## 5.1 Potential Criticisms

An obvious criticism for the model seems to raise itself. It implies that the universe can somehow anticipate future or past events in exact detail, and then play them over at will. At first glance, this just seems too far-fetched. How could it possibly *know*? With a little more thought, however, one recognizes that such a contention would assume that there actually was a differentiation between past and future events in the universe. With this model, it is clear there would not be. Events could neither be said to be in the future or past of one another; they would just *be*. Moreover, as there is nothing to make one time (as indicated by a clock) any more special than another, there is no present moment gradually unfolding; all different events and times share equal reality (in respect to time, none except for the interval used as the reference). Although physical continuity (i.e. the capability for events to be continuous), and as such, the capability for motion and of clocks and rulers to represent intervals, would stop them from all happening at once (and to happen at all), all events and times in the universe would already be mapped out. As such, as long as it still obeyed all of its own physical laws, the universe would be free and able to play any order of events it wished. Please note that this timeless picture of reality is actually the same as that provided by relativity and the "block" universe model [26], the formalized view of space-time resulting from the lack of a "preferred" present moment in Einstein's relativity theories, in which all times and events in the universe – past, present and future – are all mapped out together, fixed, and share equal status.

## 5.2 Kaon Time Reversal Invariance Violation

Finally, another problem with this model seems to present itself. The time reversal invariance (T) violation of the decay of the weak interacting sub-atomic particle, the neutral kaon (or K-meson). While avoiding a detailed discussion of kaon decay, the rate at which neutral kaons turn into neutral anti-kaons does not exactly match the rate at which neutral anti-kaons turn into neutral kaons.[9] This not only violates CP invariance (the conjugating of all charges and inversion of parity and all spatial coordinates), but T also. When the kaon's asymmetrical behaviour was first discovered in 1964 by Christensen, Cronin, Fitch and Turlay [27], it came as a tremendous shock, as not only was CP invariance thought to hold universally, the derivatives (i.e. momenta and angular momenta) of all particle interactions were also thought to be perfectly invariant and conserved under time reversal (this being the formal definition of T). CP violation means non-conservation of T, provided that the general CPT theorem is valid – something there is no reason to doubt [28]. Such a T-violation would represent a very serious problem for the present model, as, in light of kaons, the time reversal of B→C could not be truly symmetric.

---

[9] For a good explanation of kaon decay, see, for example, Davies, P. C. W., About Time: Einstein's Unfinished Revolution. pp. 208, Viking, London, 1995.

What makes the behaviour of kaons so strange is that they seem able to tell the difference, in a local sense, between the two directions of time – past and future – showing a slight bias towards the positive time direction, or future [29, 30]. The most convincing theories addressing the observed dominance of matter over antimatter in the universe are based on this observation, with a small amount of time reversal violation being said to occur in the decay of certain heavy particles in the very early universe [28]. It has been argued that the kaon's apparent ability to discern future from past could be the result of the expansion of the universe denoting a temporal direction away from the big bang towards the future, and that the kaon is somehow able to pick up on this [12]. What exactly would it pick up on, however? In this sense, in 1970, Yuval Ne'eman, a founder of the quark theory of matter, argued that the direction of time attached to kaon decay was directly related to the universe's large-scale evolution [31]. If the universe was contracting rather than expanding, the time asymmetry of kaon decay would be reversed: a contracting matter universe would be the same as an expanding antimatter universe. The key to Ne'eman's argument was that a kaon would be able tell the difference between an expanding and contracting universe. As gravity, the relevant factor here, is currently attractive, however, and should remain so if the universe begins to contract, it is difficult to see what could trigger the kaon's reversal.

Considering the possible relation between gravity and kaon T-violation, Paul Davies asks, "It was Einstein's theory of gravitation that gave us the possibility of an expanding universe. Perhaps there is some ill-understood aspect of gravity that relates to T-violation?" [12]. He continues, "[Roger] Penrose believes that gravity holds the key to the arrow of time: that there is an intrinsic lopsidedness to time when it comes to gravitational fields. At least there is when those fields are situated in the vicinity of space-time singularities…Penrose concedes he doesn't know the origin of this lopsidedness, but thinks it might somehow link up with the kaons' T violation." Investigating such issues, in a series of papers [32, 33, 34], Gabriel Chardin found that the CP violation observed in the neutral kaon system might be well explained by "antigravity," showing "just the amount of anomalous regeneration that we call CP violation" [32], while also examining and dispelling a number of possible problems sometimes associated with such an idea. In relation to the present model, I feel the fore-mentioned considerations to be suggestive. If the kaon's ability to distinguish past from the future (in a local sense) really is related to gravity (or perhaps even to anti-gravity), something denoted by its attractiveness (or in the case of antigravity, repulsion), it should reverse under a reversal of the direction of gravity, as would be the case in this model between C→B. As such, no kaon T-violation would occur under the time reversal of B→C. Indeed, somewhat surprisingly, if gravity *is* directly linked to kaon T-violation, such a violation would actually argument this model, fitting neatly with it, while the model clearly arguments the case of kaon T-violation being connected to gravity. With the B-meson having very recently been found to also violate CP symmetry [35], the same considerations outlined here in respect to gravity, the kaon, and it seemingly being able to distinguish past from future, would presumably also apply to the B-meson, as they would to any particle that violated T symmetry.

# 6. Black Holes

In 1965, Roger Penrose proved, as he did with Stephen Hawking five-years later in relation to the big bang (and a possible big crunch) [36], that if Einstein's general theory of relativity is correct, and some specific criteria are met, a singularity must result inside a black hole [37]. This means that gravity has become infinitely strong at its centre, also causing the geometry of space-time to be infinitely curved. It is not difficult, however, to see the similarity between a black hole and a big crunch. The situation is much the same, with matter being contracted by immense gravity, becoming more and more dense and hot, and the geometry of space-time contracting towards an infinite point. However, there is a difference between the two. A black hole has the vast universe outside it, while with a big crunch, there is nothing outside the collapsing region as it represents the entire universe.

As, locally, they represent the same situations, if the second law of thermodynamics *were to* be breached as the universe contracted towards a big crunch, one should expect the same to happen as matter approached a black hole singularity. In exactly the same way as the order of events would reverse just before a big crunch so that entropy could still increase, the same should happen in a black hole. Events would immediately take up from the point where the singularity would have been had events not reversed, and in this direction, as with the big crunch, no singularity would be encountered. Furthermore, in this direction, events would be going in the direction in which entropy was still increasing and the second law would still hold. It is also interesting to note why Penrose's black hole singularity theorem would not apply here, as it assumes that all in-falling matter would converge in a regular way towards the singularity [37]. As we can see with the model being discussed, this would not actually happen. Matter would always be moving away from a (would be) black hole singularity, so none would be encountered. In relation to matter always moving away from a singularity, and as a result, one not being encountered, the same can be applied to Penrose and Hawking's big bang (and big crunch) singularity theorem [36].

As with the big crunch scenario, this situation inside a black hole could last only so long. Very quickly, events would reach the point at which the second law of thermodynamics would have been breached had events not reversed. In the current direction, events (inside the reversed region) would face an order of events (that of matter falling into the black hole and of the entire outside universe) in which, in this direction, entropy would be decreasing. This would present an interesting situation. Which order of events would take precedence? Would this force the reversal of the order of events of the whole universe back to the big bang like in the big crunch scenario, or would there be a difference between the two situations? As outlined earlier, there is a difference between them. The reversed region in the black hole would be surrounded by a much larger and more dominant system, the universe itself, while the reversed big crunch would represent the entire universe, with nothing outside of it. This would suggest that as events reached this point inside the black hole, because the region outside of the reversed region would easily be the dominant system, the order of events of the universe outside would take precedence and not be reversed. As such, this would also mean that no matter would be able to escape beyond this point from inside the reversed region of the black hole, but matter could still enter from the outside. As soon as it entered, however, it would immediately be reversed; instead of contracting towards a singularity, it would be moving away from it.

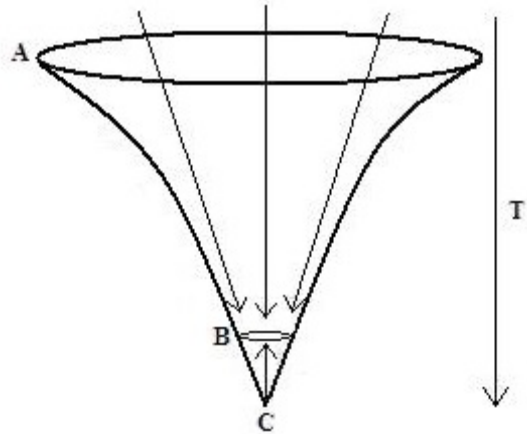

[Figure 2]

**A→B.** Matter falls into black hole, beyond its event horizon.
**B.** Point at which, as matter approaches hypothetical black hole singularity (C), the second law of thermodynamics would be breached in direction B→C.
**C→B.** Order of events reverse so that entropy can continue increasing. Matter moves outwards from where black hole singularity would have been had events not reversed, with no singularity being encountered.
**C→B.** Matter reaches original point at which the second law of thermodynamics would have been breached had events not reversed (B), and faces order of events, B→A, with entropy decreasing in this direction. As the outer region is the dominant system, order of events B→C reverses again.

It is interesting to note that once at the end of its initial reversal, matter within this reversed region could not merge and build up over time with other matter that later fell into it. This is a subtle point, and the following will hopefully help to illustrate it. If one imagines some matter falling into a black hole and reaching B at $t$ = 9.9 seconds, C being at $t$ = 10 seconds, and C→B repeating as the matter is continually reversed, one can see that it is the same epoch that repeats i.e., the interval between 10 and 9.9 seconds. As matter falling into the reversed region after that 0.1 second interval had passed, could not possibly merge with the original matter (as that epoch would now be repeating), it would have a later epoch that repeated (for example, the interval between 10.1 and 10 seconds). As such, the two epochs would be entirely causally disconnected, and the matter within them could not merge and accumulate. Indeed, in a sense, each epoch would represent its own (very small) universe – a system entirely cut off from the previous one, and which, to an outside observer, would appear to repeat endlessly, but that in respect to time, would actually be non-cyclic and happen only once. This, of course, would also mean that if and when a black hole eventually evaporated by Hawking radiation [38] (the timeframe being given by $10^{71}\ M^3$ seconds, where $M$ is solar masses), solely the matter that had not yet fallen beyond its reversed region would be returned to the outside universe. Furthermore, as the outside universe and the reversed region of a black hole would be entirely causally disconnected, although gravity would be repulsive in the reversed region, presumably no gravitational effects of the reversed region (whether attractive or repulsive) would be felt by the outside universe. With this being the case, all new matter falling beyond the reversed region would do so as if to be the *initial* cause of a singularity.

That different epochs such as this could co-exist, but not come in contact, may sound strange, but it really should not. The overall model that I have been discussing states that the same applies to the entire universe, as indeed does the relativistic "block" universe model. The only difference presently is that we are now talking about entirely disconnected systems, or if one prefers, universes. It is also interesting to note that such a reversed epoch of a black hole would very closely resemble a "baby universe" [39], offering the same solution to the black hole information loss paradox [40, 41] originally put forward by Zeldovich [42] and Hawking [43, 44], while in the sense that matter would always be moving away from a (hypothetical) singularity rather than towards it, also not be dissimilar to the Schwarzschild solution of a "white-hole" [45].

Finally, it is also interesting to see that, in the case of a big crunch, if there possibly *were* still a region outside of the reversed region due to the universe's matter not all converging in unison and some arriving later – and this outer region was the larger and more dominant system – the big crunch would be no different from a black hole. Matter within the reversed region could not escape and would continually be reversed. As further matter continued to fall into the reversed region, however, and the remaining matter in the universe had all but disappeared into it, the reversed region would eventually become the dominant system, and at the point at which it did, the order of events stretching back to the big bang would be reversed, as outlined earlier.

## 7. Predictions and Discussion

Let us now return to the overall model being discussed, both in connection to the big bang, big crunch, and black holes. As a theory, what potentially verifiable or falsifiable predictions can it make? (If, due to their nature, "potentially" is very much the operative word in some of the following cases).

I. It predicts that the universe is closed and will eventually collapse towards a big crunch. As such, it also predicts $Omega_M = > 1$.

II. It predicts that the basic big bang model [46], where a tremendously hot explosion is followed by the universe's expansion, is correct.

III. It predicts that singularities are never encountered in nature.

IV. It predicts that the order of a series of events can be reversed in certain extreme conditions (i.e. at a big crunch, inside a black hole). As such, it also predicts that the balance of heat of a closed system can never spontaneously flow from hot to hotter (i.e. instead, events would simply reverse).

V. It predicts that the universe had no beginning, has no ending, but yet is finite – in a similar sense as Hawking held that imaginary time would enable a closed universe to have no boundaries [11, 47]. As Hawking put it, the boundary condition of the universe is that it has no boundary.

VI. It predicts that time travel is not possible.

It must be said that the model's value is greatly lessened by the speculative nature of how the flow of heat from hot to hotter would cause events to reverse direction. At this point, this is as far as I am able to take it. At the same time, however, I would suggest that the overall and basic picture of the universe this model provides, makes too much sense, and fits too well, for it to be wrong, even if the detail of how the reversal is caused is incorrect. Hopefully this may also spur some theorists into attempting to elucidate further in respect to this mechanism themselves. Another big shortcoming of the model is that it is reliant on the universe being closed. At this stage, it is far from certain whether the universe really will one day stop expanding and begin to contract, or if it will continue to expand forever. Indeed, that the universe's expansion appears to be accelerating, rather than decelerating as was previously thought, seems to have generally swung the consensus towards a universe that will expand indefinitely. As discussed earlier, however, and although this model does not shed any light on the possible reason(s) behind the current accelerating expansion, or the nature of "dark energy" (or, other than the potential of such energy inevitably running down over time, how the transition to a deceleration may occur), I feel that there are some good reasons to believe that the universe eventually will collapse. This model represents a further one. Indeed, I feel that once this model has been properly understood, it becomes difficult to see how the universe could be any other way. The greatest problem with this model, however, and it applies to all theories that relate to the very early and late universe or inside black holes, is that we will never be completely certain if it is correct. We were not, and will not, be around to find out. Then again, considering what the universe would be like just after the big bang or as it neared a big crunch, this is probably just as well. Still, and not withstanding such problems, I think there is reason to justify one feeling at least modestly confident in the overall picture of the universe that this model provides. Let us go over some of the reasons why.

1. It would appear to be the only possible self-consistent cosmological model that would enable the universe to not need a beginning, not have an ending, and yet be finite in both space and time, avoid Kant's paradox, and have no paradoxical infinite values.

2. It is consistent with standard big bang cosmology.

3. It illustrates why singularities would never actually be encountered. Indeed, the reversal of an order of events just before a singularity inevitably results, would suggest itself as being nature's way of avoiding them.

4. It would appear to be compatible with general relativity, while augmenting the block universe model.

5. It does not breach any conservation laws.

6. It is without generic thermodynamic "arrow of time" reversal contradictions (i.e. with entropy decreasing). Indeed, the reversal of the order of events so that heat can still flow to cold, avoids them and guarantees that events are always causally related (irrespective of "cause" and "effect" being equally exchangeable for one another).

7. It renders meaningless the question of whether events are determined by initial or final conditions [48]. In this model, initial and final, past and future, can equally be exchanged, and they are not objectively differentiable. Other than perhaps employing basic anthropic-like reasoning such as positing that if were different, the universe would not exist (and neither would we to ask questions about it), conditions in the universe just *are*. For example, problems such as why the universe has a low entropy past [2, 3], or conditions at the big bang appear to be so "special" [4], require no *causal* explanation.

8. As we know that all of the laws of physics (except for the second law of thermodynamics) are time reversal invariant, it does not necessitate any revision of those laws, nor, as they are also time reversal invariant, does it require revision of equations such as Einstein's, Maxwell's and Schrödinger's.

9. It precludes time travel.

10. It provides insights into the possible relationship between kaon T-violation and gravity.

11. It provides a possible solution to the black hole information-loss paradox

12. Although highly counter-intuitive, once understood, it makes a great deal of sense.

## 8. Conclusion

Before finishing, I would like to make some personal comments about this model. When I initially came across it in 1998, I did nothing to pursue it further. There are at least three reasons for this. As outlined earlier, regardless of how confident I might be about the basic ideas behind the model, there is no way we will ever be able to confirm if it is actually correct. This consideration took the shine off it a little for me, as did the lack of quantitative argument behind how the reversal is caused. Another reason for my hesitancy in pursuing it further was due to just how counter-intuitive (and in some ways, perhaps even beyond the scope of people's regular thinking) it was. I did not think that many people would be able to understand it. Having since explained it to a number of people, however, and them evidentially being able to comprehend it, I have changed my mind about this. A third reason concerned the actual value of knowledge. Whether correct or not, what value might people knowing the implications of this model serve? Although I personally found them beneficial, I initially had doubts over how many others would find them the same. However, I have since changed my mind about this as well. I feel that, once properly understood, most people would find a knowledge of the implications of this model beneficial (and for various reasons, comforting), and if given the choice, most would choose to have that knowledge.[10]

To conclude then, based on the conjecture that rather than the second law of thermodynamics inevitably be breached as matter approaches a big crunch or a black hole singularity, the order of events should reverse, in this paper, a model of the universe which provides a resolution to the mystery of the origin of the universe, as well as to Kant's finite vs. infinite paradox, was presented. Implications of this model included that the universe has neither a beginning nor ending, but yet is finite; that singularities are never actually encountered in nature; there is no differentiation between past and future, and events are neither determined by initial or final conditions – they just *are*; problems such as why the universe has a low entropy past, or conditions at the big bang appear to be so "special," require no causal explanation; and that time travel is not possible.

---

[10] In this regard, I am referring to that if past and future are viewed in the regular fashion, with a meaningful differentiation being made between them, the universe can be interpreted as repeating over, exactly, an infinite number of times. While avoiding going into any detail, this has clear implications for the concepts of determinism, free will, and perhaps most significantly, the notion of life after death. There are some interesting points to be noted in connection to this last one in particular, but I feel they may be out of place in the present paper.

## 9. Acknowledgements

I would like to thank Francis Healy, Jonathan Vos Post, Philip Fellman, Paul Halpern, Roger Penrose, Brooke Yigitoz, Jojanneke van der Toorn, Andrei Linde and Lee Smolin for valuable comments and suggestions regarding the contents of this paper.